\documentstyle[11pt,newpasp,twoside,epsf]{article}
\def\src {RX\,J0806.3+1527}
\def\srcm {RX\,J1914.4+2456}

\newcommand{\AXAF}{{\em Chandra}}
\newcommand{\R}{{\em ROSAT}}
\newcommand{\A}{{\em ASCA}}

\def\edcomment#1{\iffalse\marginpar{\raggedright\sl#1\/}\else\relax\fi}
\reversemarginpar

\begin{document}
\title{Unveiling the nature of the 321s Orbital Period X--ray source \src}
 \author{GianLuca Israel, Luigi Stella}
\affil{INAF - Osservatorio Astronomico di Roma, Via Frascati 33, Monteporzio Catone, Italy}
\author{Stefano Covino, Sergio Campana}
\affil{INAF - Osservatorio Astronomico di Brera, Via Bianchi 46, Merate, Italy}
\author{Gianni Marconi}
\affil{European Southern Observatory, Casilla 19001, Santiago, Chile}
\author{Christopher W. Mauche}
\affil{Lawrence Livermore National Laboratory, 7000 East Avenue, Livermore, USA}
\author{Sandro Mereghetti}
\affil{Istituto di Fisica Cosmica G. Occhialini, CNR, Via Bassini 15, Milano, Italy}
\author{Ignacio Negueruela}
\affil{Dpto. de F\'{\i}sica, Ingenier\'{\i}a de Sistemas y 
Teor\'{\i}a de la Se\~{n}ales, Universidad de Alicante, Apdo. de Correos 99, Alicante, Spain}

\begin{abstract}
A nearly simultaneous X-ray/optical (\AXAF\ and VLT) observational
campaign of \src\ has been carried out during 2001.  These
observations allowed us to phase the X--ray and optical light curves
for the first time. We measured a phase--shift of $\sim$0.5, in good
agreement with the presence of two distinct emission regions and with
the X--ray irradiation process predictions. The \AXAF\ data allowed us
also to study in details the X-ray spectrum of \src, which is
consistent with a soft (kT$<$70\,eV) and small (R$_{BB}<$20\,km)
blackbody component.  We discuss the present findings on the light
of the models proposed so far to account for the X--ray emission
detected from \src, and its twin source \srcm.
\end{abstract}

\section{Introduction}
\src\ was discovered in 1990 by the \R\ satellite during the All
Sky Survey (RASS; Beuermann et al. 1999). However, it was only in 1999
that a periodic signal at 321\,s was detected with \R\ HRI in its soft
X--ray flux (Israel et al. 1999, hereafter I99; the discovery of
X--ray pulsations were also reported independently by Burwitz \&
Reinsch 2001).  Based on the large pulsed fraction ($\sim$100\%),
relatively low 0.5--2.0 keV flux (3.0--5.0\,$\times$\,10$^{-12}$ erg
cm$^{-2}$\,s$^{-1}$), modest distance (edge of the Galaxy is at
$\leq$1\,kpc in the direction of the source) and presence of a faint
(B=20.7) blue object in the Digitized Sky Survey 1.5'' away from the
nominal X--ray position, the source was tentatively classified as a
cataclysmic variable of the intermediate polar class (I99).

Subsequent deeper optical studies carried out during 1999-2001 both at
the Very Large Telescope (VLT; Cerro Paranal) and at the Telescopio
Nazionale Galileo (TNG; La Palma) allowed us to unambiguously identify
the optical counterpart of \src, a blue V=21.1 (B=20.7) star
consistent with that proposed by I99 and Burwitz \& Reinsch (2001),
and with no significant proper motion (Israel et al. 2002a, Israel et
al. 2002b, hereafter I02). B, V and R time--resolved photometry
revealed the presence of $\sim$15\% pulsations at the $\sim$321\,s
X--ray period (Israel et al. 2002a; I02). Independently, the discovery
of the optical counterpart was reported by Ramsay et al. (2002) ten
days later based on photometric observations carried out at the Nordic
Telescope (NOT; La Palma).

However, one of the most important piece of information was obtained
based on medium--resolution spectroscopy (VLT) of this faint object
and reported in I02.  The spectral study revealed a blue continuum
with no intrinsic absorption lines. Broad ($FWHM$$\sim$1500
km\,s$^{-1}$), low equivalent width (EW$\sim -2\div-6$\AA) emission
lines from the HeII Pickering series (plus additional emission lines
likely associated with HeI, CIII, NIII, etc.) were instead
detected. These findings, together with the period stability and
absence of any additional modulation in the 1\,min$ - $5\,hr period
range, are interpreted in terms of a double degenerate He--rich binary
(similar to the AM CVn class) with an orbital period of 321\,s, the
shortest ever recorded (I02; see also Ramsay et al. 2002).
\begin{figure}
\plotone{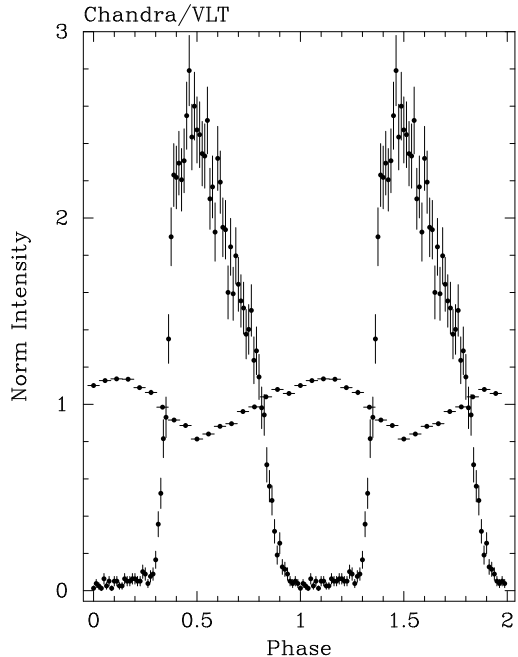}
%\plottwo{}{}
\caption{\AXAF\ ACIS--S folded light curve (nearly 100\% pulsed
fraction; eclipse is evident at phases 0.0-0.3), with superposed the
VLT optical one (R band). Optical and X--ray peaks are phase--shifted
of about 0.5.}
\end{figure}

\section{Models}
The nature of the X--ray emission detected from \src\ and its twin
source \srcm\ is still under debate.  A number of models have been
proposed in the last years (for details see Cropper et al. in
these proceedings). Among these is the double degenerate binary system
with mass transfer model which has been proposed in two flavors: with
a weakly magnetic primary (Polar--like; Cropper et al. 1998) and with
a non--magnetic accretor (Algol--like, also known as direct impact
accretion model; Marsh \& Steeghs 2002).  An additional model involves
a secondary star which does not fill its Roche lobe and crossing 
the magnetic field of the primary produces an induced electric field
(Wu et al 2002).  Finally, the possibility that \src\ and \srcm\ are
stream--fed intermediate polars (IP) seen face--on has been recently
proposed under the hypothesis that the even terms of the HeII
Pickering series might be partially due to the presence of H. In the
IP scenario the 321\,s pulsation would represents the spin period of
the accreting white dwarf (Norton et al. 2002).

The lack of good quality (and statistics) X--ray spectra has been the
most important limiting factor in the study and understanding of the
nature of this source. Moreover, the study of the possible presence of
delays of the minima and/or maxima in the pulse shapes between the
X--ray and the optical band is important for understanding the
emission geometry from \src. Therefore, the optical observations of
\src\ we requested and carried out in parallel with the X-ray \AXAF\
pointing offered us a nice opportunity to shed light on the puzzling
nature of this object.

\section{Optical/X--ray observational campaign}
On 11 November 2001 a 20\,ks \AXAF\ ACIS--S observation was carried
out as part of the AO3. At the same time we requested a 3\,hr long
Discretionary Director Time (DDT) optical observations at the 8.2\,m
VLT--U4 Yepun (the name of Venus in the old Mapuche language),
partially overlapping in time the \AXAF\ data. The data were reduced
with standard tools both for X--ray (CIAO package) and optical
observations (MIDAS and IRAF). We used the \AXAF\ observation (longer
and with higher statistics than the VLT one) to rely upon an accurate
value of the 321\,s modulation.  This is 321.51$\pm$0.03\,s (90\%
confidence level). Then, the optical and X--ray light curves were
folded at this period and the result is shown in Figure 1. A $\sim$0.5
phase--shift between the optical and X--ray peak emission is evident,
in good agreement with the predictions of the X--ray
irradiation. Specifically, we note that the minimum of the optical and
the maximum of the X--ray are at the same phase. The pulsed fraction
of the 321\,s modulation is 100\% and 14\% in the X--ray and optical
bands, respectively (consistent with results reported by I99, I02 and
Ramsay et al. 2002).
\begin{figure}
\plotone{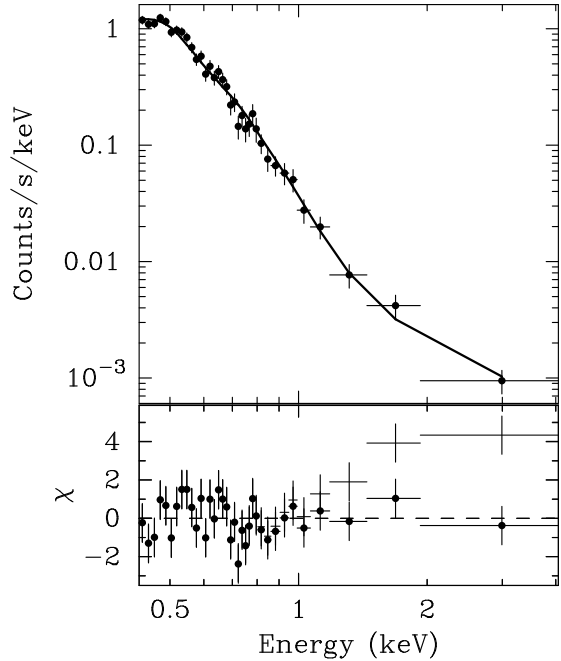}
%\plottwo{}{}
\caption{The \AXAF\ ACIS--S 20\,ks phase--averaged spectrum of \src\
fitted with a simple blackbody model. Residuals (lower panel) are
obtained for the 'raw' blackbody model (crosses) and taking into
account the pile--up (circled crosses).}
\end{figure}

The X--ray data were also used to perform a detailed spectral analysis
and below we report some preliminary results.  In particular, we
checked the double degenerate model predictions, where a hot spot (a
soft blackbody as a first approximation) is thought to be formed at the
polar caps or at the equator of the primary star: this would be
responsible for the irradiation of part of the companion star and of
the accretion stream (if any).  A first attempt in fitting the data
with a simple blackbody model resulted in the possible presence of a
high energy tail (above $\sim$2\,keV; crosses in the lower panel of
Figure 2). However, an inspection of the \AXAF\ light curve count rate
is consistent with the source being piled--up in the phase interval
0.35--0.85 of Figure 1.  Pile--up is inherently a nonlinear process
occurring in single photon counting CCD cameras. Hence, corrections
are not easy to be taken fully into account, and a parametrization of
the effects must be considered (Davis 2001).
 
With this in mind, we performed a pulse phase spectroscopic (PPS)
analysis adding to the blackbody component an ad hoc model (developped at
the Chandra Science Data Center) for the pile--up. The results of this
study are reported in Figure\,3 and can be summarised as follows: the
migration grade parameter $\alpha$ is extremely variable and confirmed
that the pile--up plays an increasingly important role as the X--ray
emission approaches the peak of the modulation, where 90\% of the
photons are affected. It is also worth noting that the size of the
corresponding blackbody is in the 10--20\,km range (assuming a
distance of 500\,pc), corresponding to an extremely small surface
fraction of a white dwarf. Moreover, the unabsorbed 0.1--2.5\,keV flux
during the pulse--on of \src\ is $\sim$3$\times$10$^{-10}$\,erg
cm$^{-2}$ $s^{-1}$ corresponding to a maximum luminosity of less than
10$^{35}$ erg $s^{-1}$ (assuming a maximum distance of 1\,kpc). There
is a factor of more than 10$^4$ between the on-- and off--pulse
fluxes.
\begin{figure}
\plotone{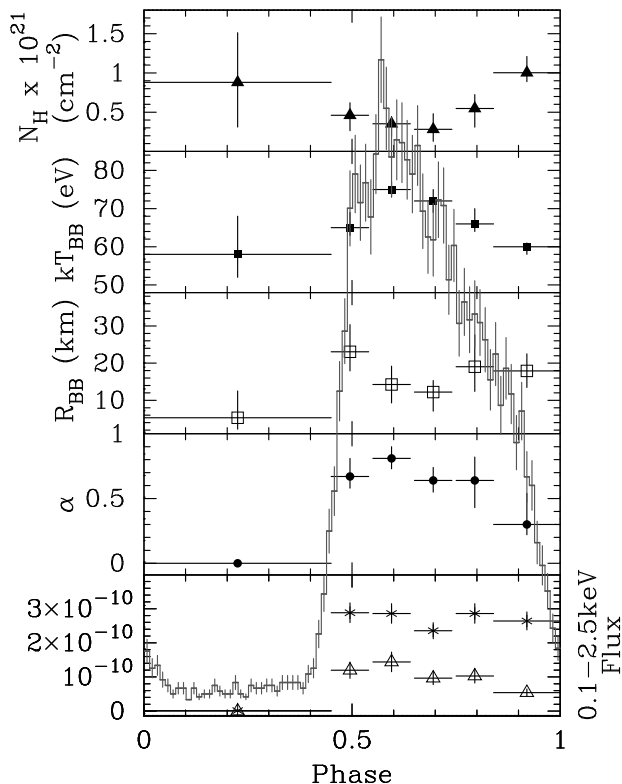}
%\plottwo{}{}
\caption{The results of the PPS analysis are reported for a number of
spectral parameters: absorption, blackbody temperature, blackbody
radius, photon migration grade fraction, and absorbed (triangles) and
unabsorbed (asterisks) flux. Superposed is the folded light curve.}
\end{figure}

Based on the above results we fitted the phase--averaged spectrum with
weighted PPS parameters. We obtained a satisfactory $\chi^{2}_{\nu}$
of 1.02 for 32 degree of freedom (see Figure 2 and the circled crosses
in the lower panel) for a blackbody with kT$\sim$65\,eV, and
absorption of $\sim$4$\times$10$^{20}$ cm$^{-2}$.  Similar results,
although with larger uncertainties, where obtained for an annular
region around the position of the source, where the pile--up is
negligible.

We also performed a radial profile study of \src\ thanks to the
spatial resolution offered by \AXAF: this was found to be in good
agreement with the expected point spread function.

\begin{figure}
\plotone{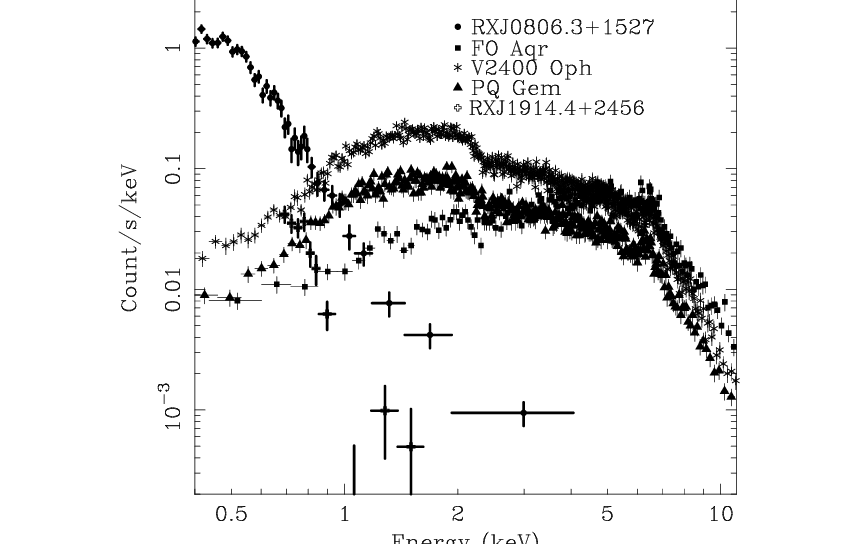}
%\plottwo{}{}
\caption{\A\ and \AXAF\ spectral comparison between \src\ and \srcm,
and a number of IPs. See the text for more details.}
\end{figure}

\section{Discussion and forthcoming observations}

The nearly--simultaneous optical and X--ray observations of \src\
carried out during 2001 allowed us have a closer look at this source.
The following preliminary implications can be reported.

(i) The nearly half pulse phase-shift between the two datasets is a
clear signature of the fact that the optical and X--ray photons come
from two different regions. This behavior is often observed in binary
systems where X--ray irradiation occurs. It is also worth noting that
the same level of phase--shift was measured for the twin source \srcm\
(Ramsay et al. 2000; this is also true for the shape of the X--ray and
optical modulations). In the Algol--like model this is easily
explained in terms of an elongated equatorial X--ray emitting region
self--eclipsed by the accreting white dwarf. In the Polar--like and
unipolar inductor models the X--ray emitting region is at the polar
cap(s) of the accretor. In both cases a 40-50\% of phase--shift is
expected.

(ii) The inferred size of the blackbody component is $\sim$20\,km, a
factor of about 10 lower than expected in the case of the electric
star model (Wu et al. 2002), and measured in IPs. However, we note
that the predicted radius in the unipolar inductor model depends on
a number of parameters; a more accurate prediction will be likely
available soon.

(iii) In order to check the IP scenario we compared the spectral
X--ray emissions of \src\ and \srcm\ with a number of IP which are
thought to be stream--fed or at least related to (FO Aqr, V2400 Oph
and PQ Gem). Figure 4 shows the result of the analysis: all the data
are taken from the \A\ database but \src\ (for which we used the
\AXAF\ data; note that due to the different energy responses of \A\
and \AXAF, only the 0.7--5\,keV energy interval can be compared). It
is evident how the \src\ and \srcm\ spectra stand out. It is difficult
to reconcile such a difference only to a different viewing angle.

(iv) Regardless the nature of \src\ the number of similarities shared
with \srcm\ is so large that we are necessarily forced in considering
them as members of the same class (currently made up of two objects).

Finally, within one year we will rely upon new observations which will
allow us to likely distinguish among the proposed scenarios.
Specifically, we will be able to detect any period derivative larger
than about 10$^{-5}$ s yr$^{-1}$. On longer time--scales the LISA
mission will search for gravitational waves from \src, the strain
amplitude of which are among the highest expected from known binary
sources (see Phinney et al. these proceedings).

\section{Acknowledgements}

GianLuca Israel is deeply grateful to Paul Plucinsky of the Chandra Team and to 
Guenther Hasinger, Piero Rosati, Roberto Gilmozzi and Martino Romaniello 
for their help in the planning end execution of the observations.

\end{document}